\documentclass[twocolumn]{aastex631}

\newcommand{\nup}{ \nu_{\rm peak} }
\newcommand{\dnup}{ \Delta \nu_{\rm peak} }
\newcommand{\dt}{ \Delta t }
\newcommand{\M}{ {\cal M} }
\newcommand{\Mori}{ {\cal M}_{\mathrm{original}} }
\newcommand{\Mshu}{ {\cal M}_{\mathrm{shuffle}} }
\newcommand{\beqr}{\begin{eqnarray} \nonumber}
\newcommand{\eeqr}{\end{eqnarray}}
\def\beq{\begin{equation}}
\def\eeq{\end{equation}}
\def\beqn{\begin{eqnarray}}
\def\eeqn{\end{eqnarray}}

\usepackage{ulem}
\usepackage{amsmath}
\usepackage{multirow}
\usepackage{soul}

\begin{document}

\title{Time-Frequency Correlation of Repeating Fast Radio Bursts:\\
Correlated Aftershocks Tend to Exhibit Downward Frequency Drifts
}

\correspondingauthor{Shotaro Yamasaki}
\email{shotaro.s.yamasaki@gmail.com}

\author[0000-0002-1688-8708]{Shotaro Yamasaki}
\affiliation{Department of Physics, National Chung Hsing University, 145 Xingda Rd., South Dist., Taichung 40227, Taiwan}

\author[0000-0002-6156-6783]{Tomonori Totani}
\affiliation{Department of Astronomy, School of Science, The University of Tokyo, Bunkyo-ku, Tokyo 113-0033, Japan}
\affiliation{Research
Center for the Early Universe, School of Science, The University of
Tokyo, Bunkyo-ku, Tokyo 113-0033, Japan}

\begin{abstract}
The production mechanism of fast radio bursts (FRBs)—mysterious, bright, millisecond-duration radio flashes from cosmological distances—remains unknown. Understanding potential correlations between burst occurrence times and various
burst properties may offer important clues about their origins. Among these properties, the spectral
peak frequency of an individual burst (the frequency at which its emission is strongest) is particularly important because it may encode direct
information about the physical conditions and environment at the
emission site. Analyzing over 4,000 bursts from the three most active
sources—FRB 20121102A, FRB 20201124A, and FRB 20220912A—we measure the
two-point correlation function
$\xi(\Delta t, \dnup)$ in the two-dimensional space of time
separation
$\dt$ and peak frequency shift $\dnup$ between burst pairs. We find a universal
trend of asymmetry about $\dnup$ at high statistical significance;
$\xi(\dnup)$ decreases as $\dnup$ increases from negative to positive values
in the region of short time separation ($\dt\lesssim0.3$ s), where
physically correlated aftershock events produce a strong time
correlation signal.
This indicates that aftershocks tend to exhibit systematically lower
peak frequencies than mainshocks, with this tendency becoming stronger
at shorter $\dt$.
We argue that the ``sad trombone effect''--the downward frequency
drift observed
among sub-pulses within a single event-- is not confined within a single
event but manifests as a statistical nature that
extends continuously to independent yet physically correlated
aftershocks with time separations up to  $\dt \sim 0.3$ s.  This
discovery provides new insights into underlying physical processes of
repeater FRBs.
\end{abstract}


\section{Introduction}
\label{section:intro}

Fast radio bursts (FRBs) are extragalactic transient objects detected in radio waves 
with millisecond durations \citep{Lorimer2007,Thornton2013}, and their source objects and emission mechanisms are 
largely still a mystery, though many theoretical models have been proposed (see
\citealt{Cordes_2019,platts2019living,zhang2020physical,petroff2022fast} for reviews). 
Some FRBs are known to produce recurring bursts, and these are likely to 
originate in neutron stars. In particular, magnetars (highly magnetized neutron stars,
see \citealt{kaspi2017magnetars,Enoto2019,esposito2018magnetars} for reviews) have been
considered a promising source of FRBs because of their abundant magnetic energy and 
the bursts of X-rays and gamma-rays that they occasionally induce.
In fact, on April 28, 2020, two extremely bright radio bursts 
(FRB 20200428)  were detected from the Galactic magnetar SGR 1935+2154. Although a few orders of magnitude fainter than typical extragalactic FRBs, these bursts demonstrated that magnetars can produce FRB-like events, offering evidence for FRB-magnetar connections
\citep{andersen2020bright,bochenek2020fast}. 

\begin{table*}
\footnotesize
\centering
\caption{Summary of the FRB datasets }
\begin{tabular}{lccccccccc}
\hline\hline
Source (sample)
 &Telescope& Period  & ${N_{\rm day}}^a$ & ${t_{\rm obs}}^b$ & ${N_{\rm event}}^c$ ($N_{\rm event}^{\rm all}$)        & ${r_m}^d$
 & ${\Mori}^e$ & ${\Mshu}^f$&${z_\M}^g$  \vspace{0.08cm} \\ \cline{8-10}
 Refs.& Band (GHz)& (MJD) &  &(day) & & (day$^{-1}$) &\multicolumn{3}{c}{$\dt<300{\:\rm ms}$} \\

& & &  &  &  & &  \multicolumn{3}{c}{$30{\:\rm ms}<\dt<300{\:\rm ms}$}
\vspace{0.08cm}\\
 
  \hline\hline
20201124A (X22)  &FAST& 59307.33--59360.18 &45 &3.27
&1135 (1863) 
         &210
& $-7.3$ & $0.53\pm2.4 $ & $-3.3\,\sigma$ \\
\citet{xu2022fast} & 1--1.5 & & &    & & & $-5.3$ & $0.55\pm2.3$ & $-2.5\,\sigma$ \\\hline
20201124A (Z22)  &FAST& 59482.94--59485.82 &4 &0.156
&1081 (1461) 
         &12000
& $-32$ & $0.53\pm3.6$ & $-8.9\,\sigma$ \\
\citet{Zhou2022} & 1--1.5 & & &    & & &  $-21$ & $0.67\pm 4.2$ & $-5.1\,\sigma$ \\\hline
20220912A (Z23)  & FAST & 59880.49--59935.39 & 17 & 0.31
& 983 (1076)
         & 4600
& $-11$ & $0.11\pm2.5$ & $-4.6\,\sigma$  \\
\citet{zhang2023frbs} & 1--1.5 &    & & & &  &$-15$ & $0.0\pm2.11$ &$-6.2\,\sigma$  \\\hline
20121102A (J23)  &Arecibo& 58409.35--58450.28 &8 &0.265
&895 (1027) 
                &4000
& $-13$ & $0.31\pm2.6$ & $-5.2\,\sigma$ \\
\citet{jahns2023frb}& 1.15--1.73 & &  & &   && $-4.7$ & $ 0.0\pm2.4$ & $-2.0\,\sigma$\\\hline
20121102A (J23g)$^h$  &Arecibo& 58409.35--58450.28 &8 &0.267
&753 (849) 
                &3300
                & $-3.6$ & $-0.16\pm2.1$ & $-1.6\,\sigma$ \\
\citet{jahns2023frb}& 1.15--1.73 &  &  &  &&& $-2.4$ & $0.09\pm2.0$ & $-1.2\,\sigma$  \\\hline
\hline
\end{tabular}
\raggedright \\
\vspace{0.2cm}
$^a$Total number of days with observations during which multiple bursts were detected (\S \ref{ss:method})\\
$^b$Total observation duration (\S \ref{ss:method})\\
$^c$Total number of events after applying a $30$ MHz cut at both edges of the observing band (\S \ref{s:data})\\
$^d$Mean event rate weighted by the number of bursts over all observation days (\S \ref{s:data})\\
$^e$Disparity moment $\M$ (defined by equation \ref{eq:moment}) calculated for the original $\nup$ dataset (\S \ref{ss:result})
\\
$^f$Mean ($\Mshu$) and 1-$\sigma$ standard deviation ($\sigma_{\Mshu}$) of $\M$ calculated from 200 randomly shuffled $\nup$ datasets (\S \ref{ss:result}; Figures. \ref{fig:1Dxi}--\ref{fig:1Dxi_30dt300})\\
$^g$Standard $z$-score of $\Mori$ defined by $z_\M = (\Mori-\Mshu)/\sigma_{\Mshu}$ (\S \ref{ss:result}; Figures. \ref{fig:1Dxi}--\ref{fig:1Dxi_30dt300})\\
$^h$When sub-bursts are grouped together  (\S \ref{s:data})\\
\label{table:FRB}
\end{table*}
More than several thousand FRB bursts have already been detected from 
several extragalactic FRB repeaters, and detailed statistical studies are possible.
An interesting fact already established is that the burst wait-time distribution is bimodal \citep[e.g.,][]{Li+21}.
Although the long-side peak of the bimodal distribution can be
explained by events occurring randomly by a Poisson process \citep{jahns2023frb}, the origin of the shorter peak has 
not been established \citep[e.g.,][]{wang2017sgr,wadiasingh2019repeating,Wang_2023}
.

\citet[][hereafter TT23]{Totani+23} analyzed the two-point correlation function in the two-dimensional space of occurrence time and energy of repeating FRBs, revealing that the statistical characteristics of FRBs are remarkably similar to those of earthquakes, while differing from those of solar flares. Building on TT23, \citet{tsuzuki2024} identified similarities between periodic radio pulsations from a magnetar and FRBs.

These investigations into the time-energy correlation of FRBs and magnetar radio pulses suggest the presence of a shared time correlation, specifically following the Omori-Utsu law $\xi(\dt)\propto (\dt+\tau)^{-p}/\tau^{-p}$, well-known to hold in earthquakes \citep{omori1895after,Utsu1957,Utsu1961}. Here, $\xi(\Delta t)$ is the correlation function for events with time interval $\Delta t$, and $\tau$ is the characteristic timescale, which is comparable to the typical event durations (i.e., $\sim$ ms for FRBs). 
Interestingly,  the only difference in between FRBs and earthequakes lies in the value of the Omori-Utsu index, $p$, and this unique correlation in time appears universal among different repeating FRB sources.

In contrast, burst energies in FRBs show little to no correlation, suggesting that their energy may be generated randomly. For earthquakes, however, weak correlations between time and energy have been identified through detailed statistical analyses \citep{Lippiello2008,deArcangelis2016}. These studies also reveal an asymmetry regarding the energy difference between pairs (i.e., more pairs with negative energy shift, where aftershock energy is smaller than the mainshock) in certain datasets. Although no such energy correlation or asymmetry has been observed in FRBs, it remains unclear whether this reflects an intrinsic lack of correlation or limitations due to sample size or detection sensitivity. Future studies are needed to clarify these differences (TT23).

This lack of correlation in energy raises an intriguing question about whether unique correlations can be identified in other FRB properties that may be more closely linked to the underlying radiation mechanisms. Given the unknown production mechanism of FRBs, exploring potential correlations between burst occurrence times and various burst properties may provide important insights. Among these properties, spectral peak frequency is particularly interesting, as it may encode direct information about the physical conditions and environment at the emission site (e.g., \citealt{Lyu2024}).

In this study, we perform a two-point correlation function analysis similar to TT23, replacing burst energy with burst peak frequency to examine whether there are correlations in the peak frequency of FRBs, given the established time correlation and the absence of energy correlation. This exploration aims to deepen our understanding of the mechanisms behind FRB emissions. The paper is organized as follows. We describe our FRB data set in \S \ref{s:data}. We describe our methodology in \S \ref{ss:method} and present the result in \S \ref{ss:result}. 
The implications of our findings are discussed in \S \ref{s:discussion} and we summarize and conclude in \S \ref{s:conclusion}. 

\section{Data}
\label{s:data}

We compute the correlation functions for five datasets of FRBs observed by the Arecibo and FAST telescopes from three repeating sources, as listed in Table \ref{table:FRB}. Below, we provide detailed descriptions of these FRB datasets.

{\it FRB 20201124A (X22 \& Z22)--} 
FRB 20201124A \citep{CHIME2021} is a repeater known for its high activity.
It is localized to a Milky Way-like, barred spiral galaxy
at $z=0.0979$ \citep{xu2022fast}. We take the barycentric arrival times $t$  of the X22 \citep{xu2022fast} 
from the tables in the original papers. Since peak frequency information was not provided in the original paper, we obtained the one-dimensional spectra extracted from the full dynamic spectra (as archived in \citealt{Wang2023}) for all bursts from the authors. We then fitted these spectra with a 1D Gaussian function, $\propto \exp{\left\{-(\nu-\mu)^2/(2\sigma^2)\right\}}$, using the fitted $\mu$ as the peak frequency $\nup$ \citep[e.g.,][]{aggarwal2021comprehensive,Zhou2022,zhang2023frbs}. Additionally, we obtained the observation log for the X22 data set directly from the authors. 

There was another observation of this source during a period of enhanced burst activity \citep[][hereafter Z22]{Zhou2022}, which we use to compare results from the same source at different activity levels. We took $t$
and $\nup$ (``$\nu_0$'' in the table of Z22) of all 1,461 bursts from
the original paper (Z22). 
We used the observation log from the original paper.

{\it FRB 20220912A (Z23)--} 
FRB 20220912A \citep{McKinven2022} is another repeater being located at
a position that is consistent with a host galaxy of stellar mass 
$\sim 10^{10} M_\odot$ at  $z=0.0771$ \citep{Ravi2023}.
We used barycentric times $t$  and peak frequencies $\nup$ of 1,076 bursts from the Z23 dataset \citep{zhang2023frbs} as presented in the original paper. We note that their $\nup$ values were derived using the same method (i.e., a 1D Gaussian function) that we employed in our analysis of the X22 dataset.

\begin{figure*}
\centering
\includegraphics[width=8.5cm]{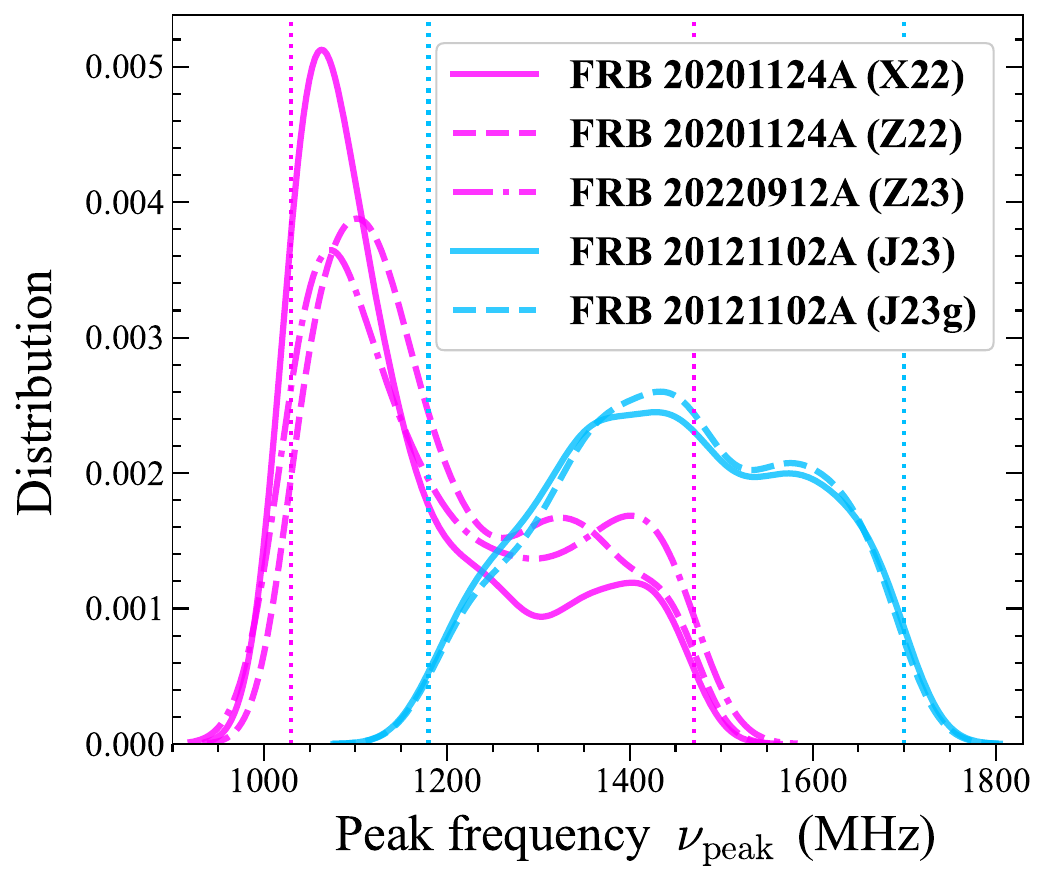}
\caption{Distribution of peak frequency data used in this study. The magenta (FAST) and light blue (Arecibo) curves represent the full-band datasets, while vertical dotted lines show the trimming condition with 30 MHz edges to minimize observational biases (see \S \ref{s:data} for the details).}
\label{fig:nu_peak_dist}
\end{figure*}

{\it FRB 20121102A (J23 \& J23g)--} 
FRB 20121102A is the first discovered \citep{Spitler2014,Spitler2016}, 
highly active, and most well-studied repeater
located in a star-forming dwarf galaxy at redshift 
$z = 0.193$ \citep{Bassa2017,Tendulkar2017}.
\citet{jahns2023frb} lists all 1,027 bursts and the 849 independent events (after grouping sub-bursts into one independent event based on their own criteria)  in their tables. We primarily use the complete sample of bursts, referred to as J23, while the grouped sample is referred to as J23g. The latter is examined to assess how the results change when events are grouped by individual observers. We take the  barycentric time $t$
and peak frequency $\nup$ of bursts (``$\nu_0$'' in the J23 table and ``$f_{\rm cent}$'' in the J23g table)  from
the original paper \citep{jahns2023frb}. 
Additionally, we used the observation log 
(start and end times of each observation) provided in \citet{jahns2023frb}.

Figure \ref{fig:nu_peak_dist} illustrates the distribution of peak frequency data used in this study. The $\nup$ distribution appears bimodal, as noted in previous studies. This pattern might result from a combination of intrinsic properties and potential observational biases, though their origins and statistical significance of this bimodality remain unclear \citep{jahns2023frb,Zhou2022,zhang2023frbs,Lyu2024}. Since this study focuses on other aspects, a detailed investigation of these observational effects is beyond its scope. Nevertheless, given that the observations were conducted with limited frequency coverage, it is important to interpret the ``peak'' frequency carefully, as the true value may lie outside the observed band \citep{Lyu2024}.
To mitigate such biases, we exclude the events falling within 30 MHz of the band edges \citep[e.g.,][]{jahns2023frb,zhang2023frbs}. This event cut reduces the sample size by $9$\% (Z23), $11$\% (J23g), $13$\% (J23), $26$\% (Z22), and $39$\% (X22), depending on the dataset (see $N_{\rm event}/N_{\rm event}^{\rm all}$ in Table \ref{table:FRB}).
The choice of frequency cutoffs can influence the selection function of $\nup$ in the entire sample. However, our analysis rather focuses on the ``differences" in $\nup$ between burst pairs ($\dnup$; see \S \ref{ss:method}), which are less sensitive to these selection effects. Consequently, the two-dimensional correlation between the time separation of burst pairs ($\dt$) and $\dnup$ remains largely unaffected.

Finally, the redshifts of these FRB sources ($z<0.2$) are small enough that cosmological effects are negligible, so no corrections for cosmological time dilation have been made to the time and frequency measurements.

\section{Correlation analysis}

\subsection{Methods of correlation function calculations}
\label{ss:method}
The method for calculating the two-point correlation function follows the approach in TT23, and here we briefly describe the difference
from the previous study.  We calculate the two-point correlation function $\xi$ in the two-dimensional space of $\dt$ and $\Delta \nup$, where $\dt \equiv t_2 - t_1$ ($t_1 < t_2$) represents the time difference between two events occurring at times $t_1$ and $t_2$, and $\Delta \nu_{\rm peak} \equiv \nu_{{\rm peak},2} - \nu_{{\rm peak},1}$ represents the difference in peak frequencies, with $\nu_{{\rm peak},1}$ and $\nu_{{\rm peak},2}$ corresponding to the events at $t_1$ and $t_2$, respectively. The correlation function $\xi (\dt, \dnup)$ quantifies the excess pair density relative to the uncorrelated case. Therefore, the number of pairs ($dN_{\rm p}$) in a bin at ($\dt$, $\dnup$) is given by:
\begin{equation}
    dN_{\rm p} = (1+\xi)\,\overline{n}_{\rm p}\, d(\dt)\, d(\dnup) \ ,
\end{equation}
where $\overline{n}_{\rm p}$ represents the expected pair number density for an uncorrelated distribution.
To estimate $\overline{n}_{\rm p}$, we generate random, uncorrelated bursts using a Monte Carlo method. 
To generate random event times, we first calculate the average observed burst rate during each continuous observation period. Using this rate, we simulate event times by generating waiting times through a Poisson process, which is appropriate for modeling random, independent events occurring at a constant average rate. For the random event peak frequencies, we empirically construct it from the data sets, assuming that the distribution remains constant during each continuous observation period. We then randomly sample from this distribution, following the methodology applied to random event energies in TT23.
To minimize statistical errors, the random sample size ($N_{\rm R}$) is set to be sufficiently larger than that of the real data ($N_{\rm D}$) for each dataset. To balance statistical convergence with computational efficiency, we generated random samples 30 times larger than the real data for datasets with over 1,000 events (X22 and Z22), and 40 times larger for those with fewer events (Z23, J23, and J23g). We confirmed that increasing the random sample size beyond these factors did not significantly alter the correlation function results, indicating that our chosen sample sizes are sufficient for reliable analysis. The correlation functions are calculated using the natural estimator \citep{Peebles1974}
\begin{equation}
    \xi(\dt, \dnup) = \frac{DD^\prime}{RR^\prime} -1\ .
\end{equation}
Here the normalized pair counts $(DD^\prime, RR^\prime)$ are related to their original pair counts $(DD, RR)$ as follows:
\beqn
    &DD& = \frac{N_{\rm D}(N_{\rm D}-1)}{2}\,DD^\prime
    \ , \nonumber\\
    &RR& = \frac{N_{\rm R}(N_{\rm R}-1)}{2}\,RR^\prime
    \ , \nonumber
\eeqn
where $DD$ and $RR$ represent the number of pairs in the real and random samples within a given bin of the $\dt$-$\dnup$ space. 
We use the natural estimator instead of the Landy-Szalay (LS) estimator \citep{Landy+1993}, as the LS estimator can yield unphysical results of $\xi < -1$ when the sample size is limited. However, we verified that using the LS estimator barely changes the results presented in this work.

In this study, Poisson errors in pair counts are used to estimate the uncertainties in the correlation function. While jackknife errors are more appropriate than Poisson for accurately accounting for the covariance between different bins, they tend to be less precise when sample sizes are small. Previous work (TT23) demonstrated that the resulting jackknife errors were not significantly different from Poisson errors.
Therefore, using Poisson errors is sufficient for this study, which does not require rigorous parameter error estimation \citep{tsuzuki2024}. Poisson statistical errors are numerically computed to take into account small-number statistics accurately \citep[e.g.,][]{Gehrels1986}.

The observation period for a single FRB dataset spans multiple days, with continuous observations typically lasting only a few hours per day (see Table \ref{table:FRB}). Following TT23, the entire observation period was divided into sub-periods, where each sub-period comprises events detected on the same observing block. Pairs of events across different observing blocks were not considered. We counted pairs for each observing block and summed them across all blocks to compute the correlation function for the dataset.
The event occurrence rate and $\nup$ distribution were held constant for random data generation within each sub-period. If there were multiple interruptions during the day's observations (e.g., X22), the correlation function was calculated without generating random events during those gaps, using information on the start and end times of each observing run. 

\subsection{Results}
\label{ss:result}

\begin{figure*}
\includegraphics[width=\textwidth]{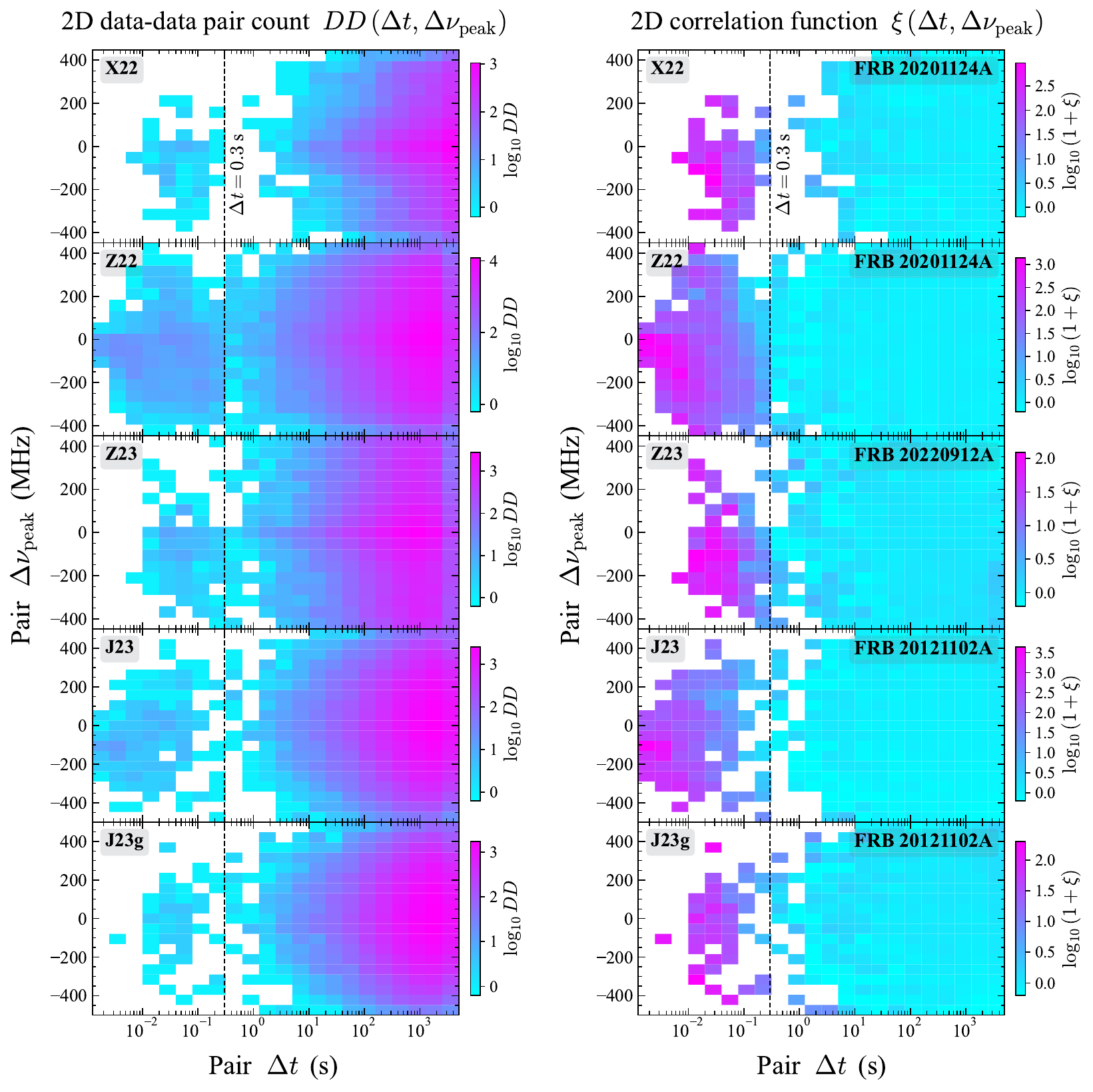}
\caption{Distribution of DD pairs (left) and two-point correlation function $\xi$ (right) in $\dt$-$\dnup$ phase space for different repeating FRB datasets. The vertical dashed black lines represent the $\dt = 0.3$ s threshold for analyzing short-time separated pairs (see \S \ref{ss:result}).
}
\label{fig:2Dxi}
\end{figure*}

Fig. \ref{fig:2Dxi} shows the two-dimensional DD pair count and
correlation function of different FRB datasets in the $\dt$-$\dnup$ space. 
A clear bimodal distribution is observed in the $\dt$ direction (left panel), with the dominant component at $\dt > 1\,\mathrm{s}$ corresponding to the longer side of the bimodal wait time distribution. The absence of a distinct correlation signal in the $\dt > 1\,\mathrm{s}$ region (right panels) suggests that these large time separation events can be described by an uncorrelated Poisson process, consistent with previous studies. In contrast, Similar to what was seen in the two-dimensional time-energy correlation function of TT23, strong correlation signals are confirmed across all datasets for $\dt < 1$ s, corresponding to the shorter side of the bimodal distribution.

Intriguingly, the two-dimensional correlation function (right panels of Fig. \ref{fig:2Dxi}) for short time-separation pairs ($\dt < 1$ s) suggests that the signal may vary in the frequency direction, implying a potential correlation among aftershocks. This differs from the trend observed in TT23 for FRB energy $E$, where the correlation function at $\dt < 1$ s was uniform in the $\Delta E$ direction. To investigate the possible frequency dependence of $\xi$ for short-time separation events, we focused on pairs with $\dt$ smaller than $\sim 0.3$ s, which serves as the rough universal boundary between short- and long-time components in the waiting time distribution (indicated by the vertical dashed lines in Fig. \ref{fig:2Dxi}). We note that this boundary is approximate, and our results are not sensitive to the exact choice of this value. We then analyzed how the correlation function $\xi$ depends on $\dnup$ within this short-time regime.

\begin{figure*}
\includegraphics[width=\textwidth]{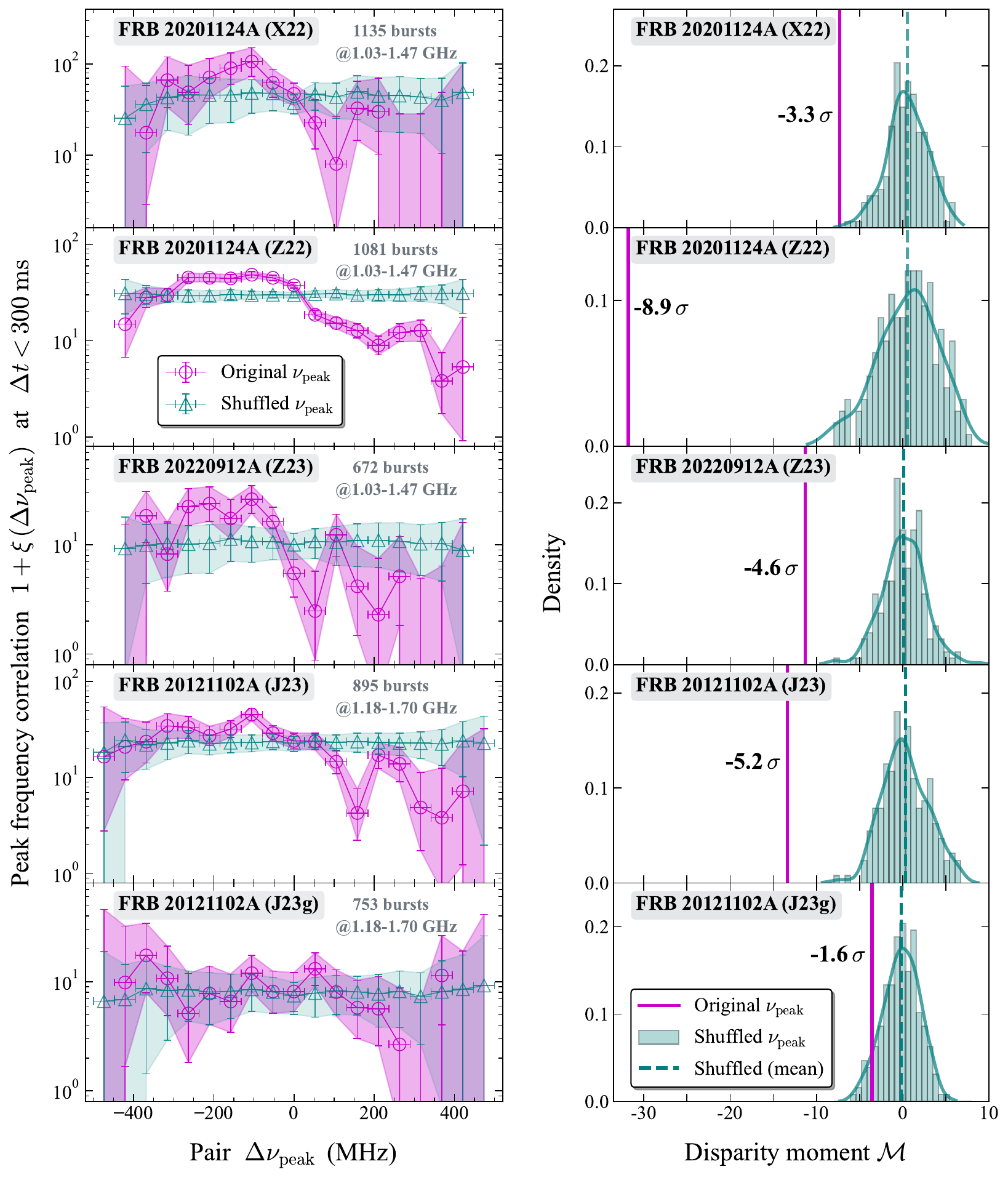}
\caption{{\it Left}: The peak frequency correlation function, $\xi(\dnup)$, for bursts with short time separation $\dt < 300$ ms from five FRB data sets. The magenta open circles represent the results for the original $\nu_{\rm peak}$ data, with shaded regions indicating uncertainties due to Poisson statistics in the DD and RR pair counts. 
The teal open triangles represent the uncorrelated $\xi$, derived from the averaged results of randomly shuffled $\nu_{\rm peak}$ data, with the shaded regions indicating the statistical scatter among the shuffled datasets. 
{\it Right}: Comparison of the $\xi(\nup)$ disparity moments, $\M$, for the original datasets with the moment distribution of the shuffled datasets. The vertical dashed lines represent the mean $\M_{\rm shuffle}$, and the standard $z$-score of $\M_{\rm original}$ is shown in each panel (see also Table \ref{table:FRB} for the details).
}
\label{fig:1Dxi}
\end{figure*}

The left panels of Fig. \ref{fig:1Dxi} show the one-dimensional correlation function $\xi(\dnup)$ for short-time separated pairs with $\dt < 0.3$ s. There is a notable $\dnup$-dependence in $\xi$, characterized by an increase in $\xi$ at negative $\dnup$ values, resulting in a clockwise rotational pattern in the $\xi(\dnup)$ plot.
To test whether there is a significant dependence between  $\dnup$ and $\xi$, we randomly shuffle the $\nup$ values within the original $(t, \nup)$ dataset before calculating the DD pairs and then perform the same analysis. This shuffling should remove any correlation between $\dnup$ and $\dt$ while preserving the overall distribution of $\nup$. By repeating this randomization over 100 times for each dataset and averaging the results, along with calculating the dispersion in $\xi$, we obtained the uncorrelated $\xi(\dnup)$ and 1-$\sigma$ statistical errors, which is also shown in the left panels of Fig. \ref{fig:1Dxi}. As expected, the $\xi$ signal for the shuffled dataset is flat across $\dnup$, in stark contrast to the $\dnup$ dependence seen in the original dataset.

\begin{figure*}
\includegraphics[width=\textwidth]{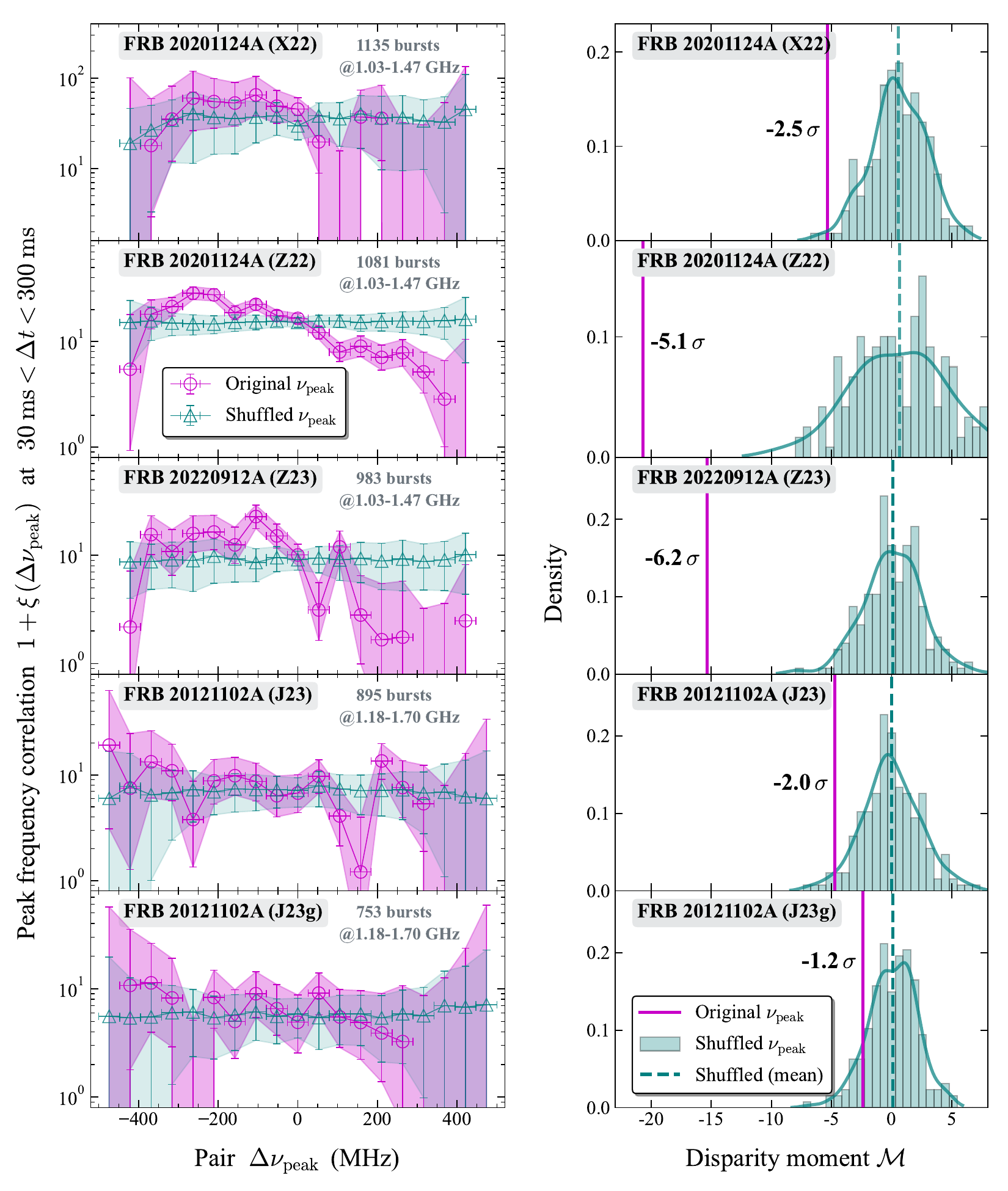}
\caption{Same as Fig. \ref{fig:1Dxi}, but for burst pairs with time separation $30$ ms $<\dt<300$ ms.
}
\label{fig:1Dxi_30dt300}
\end{figure*}

To quantify this pattern further, we introduce a disparity moment for a given correlation function $\xi(\dnup)$ by
\begin{equation}\label{eq:moment}
\M \equiv \sum\limits_{i} \,\Delta \widetilde{\nu}_{i}\,  \Delta \widetilde{\xi}_i
\, ,
\end{equation}
where $i$ denotes the $\Delta \nu_{\rm peak}$ grid, and $\Delta\widetilde{\nu}_{i}\equiv \Delta\nu_{{\rm peak},\, i}/|\Delta \nu_{\rm peak,\,{\rm max}}|$ is the $i$-th $\dnup$ normalized by the maximum $|\dnup|$ (determined by the observing band) for each dataset, and $\Delta \widetilde{\xi}_i$ represents the deviation of $i$-th bin $\xi$ from the average $\xi$ across all $\Delta \nu_{\rm peak}$ bins, weighted by the the inverse of the Poisson statistical error of $i$-th bin to reduce the influence of bins with larger errors, i.e., $\Delta \widetilde{\xi}_i=(\xi_i -\left\langle\xi\right\rangle)/\sigma_{\xi_i}$. Poisson errors are asymmetric around $\xi_i$, with upper and lower errors. However, the magnitude $\sigma_{\xi_i}$ of remains unchanged regardless of whether the upper or lower error is used. Thus, we calculate $\sigma_{\xi_i}$ by simply taking their average.
Defined in this way, a clockwise trend results in $\M$ being biased towards negative values, while an anti-clockwise trend biases $\M$ toward positive values. A value of $\M=0$ indicates that the correlation function is line-symmetric to $\dnup=0$.

The disparity moment is calculated consistently for both the original and shuffled datasets, using all the $\dnup$ bins.  The original data's moment, ${\cal M}_{\rm original}$, is then compared to the distribution of shuffled moments, ${\cal M}_{\rm shuffle}$, to assess the statistical significance of the observed correlation. The computed ${\cal M}$ results are summarized in Table \ref{table:FRB} and the right panels of Fig. \ref{fig:1Dxi}. 
As expected, the moment distribution for random dataset centers around $\M=0$ with some dispersion, while the moment for the original dataset falls within $\M<0$.
The statistical significance of the correlation being stronger at negative frequency shifts is $3.3$, $8.9$, $4.5$, and $5.3\,\sigma$ for the  X22, Z22, Z23, and J23 datasets, respectively (when these results are combined, the overall significance increases to $12\,\sigma$). This indicates that the finding is universal, regardless of the FRB source. 
We also note that two different datasets from the same source, FRB 20201124A, collected during distinct burst episodes -- active (X22, $3.3\,\sigma$) and extremely active (Z22, $8.9\,\sigma$) states -- exhibit similar results, with the latter showing stronger statistical significance. However, examining Fig. \ref{fig:2Dxi}, the shapes of $\xi$ for X22 and Z22 appear consistent, suggesting that the larger statistical errors in X22 are simply due to its smaller event sample size. This suggests that the shape of $\xi$ is independent not only of the source but also of the burst activity level.  

Furthermore, when sub-bursts are grouped by individual observers, the significance drops from $5.2$ to $1.6\,\sigma$ for the J23 $\to$ J23g dataset.  The method of grouping sub-bursts varies considerably across observers, but in general, grouping reduces the number of closely time-spaced (i.e., small $\dt$) pairs (compare J23 and J23g results on the left panels of Fig. \ref{fig:2Dxi}). This further demonstrates that the observed trend strengthens as the time separation between bursts decreases (see \S \ref{s:discussion}).

To examine this effect in all datasets more systematically, we additionally investigated the dependence of the results on $\dt$. In the above, we consider all the burst pairs including very closely spaced burst pairs with $\dt < 300$ ms in the analysis, which are often considered sub-bursts within a single burst in observations. To exclude the contribution of such very near-time pairs, we reanalyzed the data with a lower limit of $\dt$ set to 30 ms. The results are shown in Fig. \ref{fig:1Dxi_30dt300} and summarized in Table \ref{table:FRB}. The significance of the negative frequency shift remains notable even at longer timescales of $30$ ms $< \dt < 300$ ms (overall significance combined across X22, Z22, Z23 and J23 is $8.7\,\sigma$). However, compared to the case that includes shorter separation pairs ($\dt < 300$ ms, Fig. \ref{fig:1Dxi}), the overall significance slightly decreases (from $12$ to  $8.7\,\sigma$).

In summary, the results indicate a stronger tendency for aftershock $\nup$ values to decrease as $\dt$ becomes smaller, with this trend extending continuously from $\dt \sim 1$ ms to $300$ ms and becoming more pronounced at shorter $\dt$.
This behavior is already visible in the two-dimensional correlation function $\xi(\dt, \dnup)$ (right panels of Fig. \ref{fig:2Dxi}) and is statistically confirmed through the moment analysis of the one-dimensional correlation function $\xi(\dnup)$ (Figs. \ref{fig:1Dxi} and \ref{fig:1Dxi_30dt300}).

\section{Discussion}
\label{s:discussion}
\subsection{Interpretation}

We find evidence of a universal time-frequency correlation between individual burst pairs at short time separations ($\lesssim0.3$ s). Specifically, there is a stronger tendency for aftershock $\nu_{\text{peak}}$ values to decrease as $\Delta t$ shortens, with this trend spanning continuously from $\dt \sim 1$ ms to $300$ ms\footnote{In fact, $\nup$ decreases for specific pairs with $\Delta t \sim 100$ ms were also reported in some literature \citep{jahns2023frb,Zhou2022}, though without a quantitative correlation analysis. Our analysis of these datasets further corroborates this observation, providing additional insight into the underlying continuous drift phenomenon. }.
This finding is somewhat surprising because burst pairs with relatively large time separations of $\Delta t \gtrsim 100$ ms, corresponding to the smaller peak in the bimodal wait-time distribution, are generally considered independent bursts rather than multi-peaked light curves within a single event, despite exhibiting some temporal correlation (TT23).


When discussing downward frequency shifts in FRBs, the most well-known (yet still unexplained) phenomenon is the sub-burst downward frequency drift, often referred to as the ``sad trombone'' effect \citep[e.g.,][]{Hessels2019,CHIME2019drift,CHIME2020drift,Fonseca2020,Hilmarsson2021,jahns2023frb,Pastor-Marazuela2021,Platts2021,Pleunis2021,zhang2022fast}. This effect occurs within bursts containing multiple sub-bursts, where the central frequencies gradually drift to lower values over time, typically on timescales of $\Delta t \lesssim 10$ ms. 

Notably, our correlation analysis, which considers all possible pairs including consecutive bursts, has captured this effect in both the Z22 and J23 datasets (see the enhanced $\xi$ in the right panels of Fig. \ref{fig:2Dxi}) at $\Delta t \lesssim 10$ ms. Moreover, our results suggest that $\nup$ does not always decrease, but rather that there are statistically more pairs with a decreasing $\nup$ than with an increasing one. This trend becomes more pronounced at shorter $\Delta t$, meaning that for small $\Delta t$, downward $\nup$ shifts are more likely than upward shifts. 
This trend is consistent with the fact that repeating FRBs rarely show evidence of upward frequency shifts between consecutive sub-bursts at small $\Delta t$ (Z22); instead, they predominantly exhibit downward frequency shifts, characteristic of the sad-trombone effect.

 \begin{figure*}
\includegraphics[width=\textwidth]{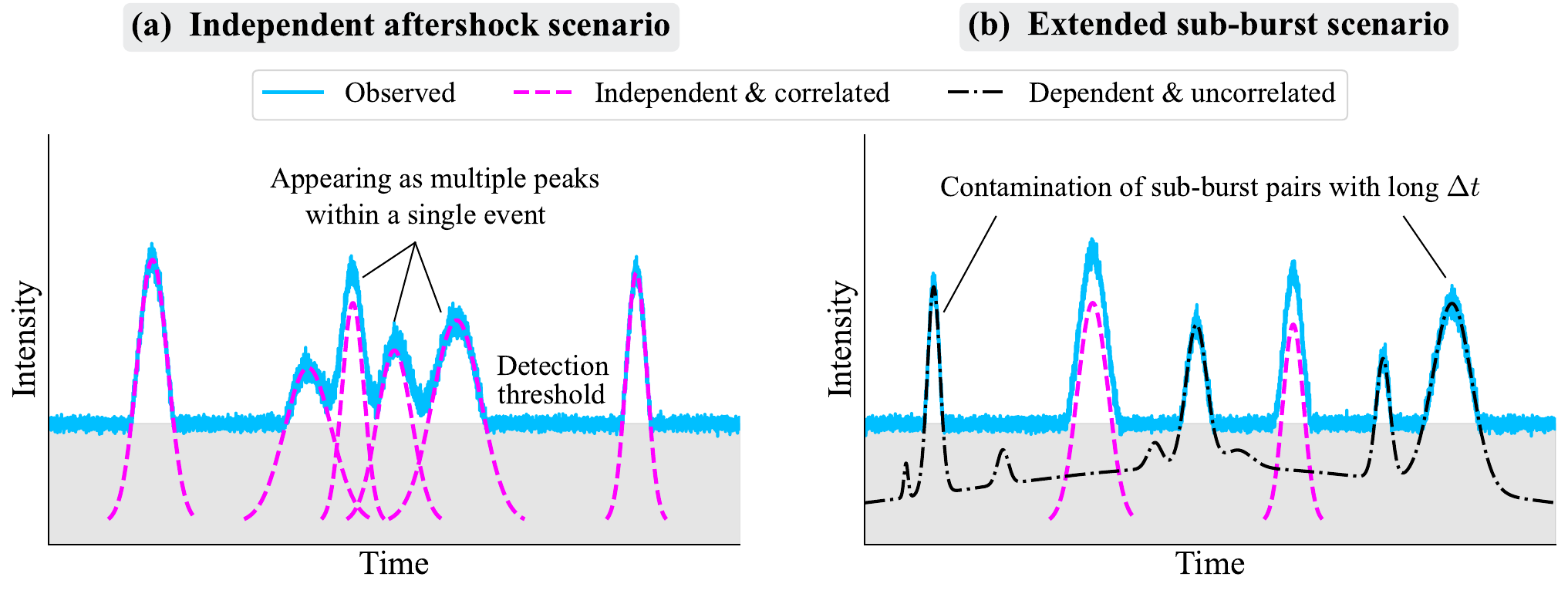}
\caption{Schematic illustration of two possible interpretations. {\it Left} (a): The independent aftershock scenario, where all bursts are independent yet physically correlated aftershocks. {\it Right} (b): The extended sub-burst scenario, which includes a mix of independent, physically correlated pairs as in scenario (a) and widely separated pairs that are actually part of the same event, with their broad component undetectable due to sensitivity limits. In both panels, the observed light curve is depicted by a solid blue line (with added noise), while independent events are represented by dashed lines. In panel (b), a single event with interdependent yet physically uncorrelated apparent substructures is represented by a dash-dotted line. The shaded gray regions indicate areas below the detection threshold.
}
\label{fig:scenarios}
\end{figure*}

To interpret these further, we now discuss the relationship between our findings and the downward frequency drift observed among sub-bursts within a single event, particularly for $\Delta t \lesssim 10$ ms. Two possible scenarios are proposed (see Fig. \ref{fig:scenarios})
:

\begin{enumerate}
    \item {\bf Independent aftershock scenario} --
    One possible scenario is that independent aftershocks, though physically correlated, generally exhibit decreasing frequencies, with this trend becoming more pronounced as $\dt$ decreases. This behavior continues smoothly over the $\sim1$--$300$ ms range. Crucially, within this framework, what are typically considered sub-bursts within a single event at $\dt \lesssim {\cal O}(10)$ ms may be aftershocks occurring in quick succession. If a subsequent independent event occurs and its flux rises before the flux of the previous burst falls below the detection threshold, it becomes fundamentally indistinguishable from multiple sub-bursts within a single burst (see the left panel of Fig. \ref{fig:scenarios}). This implies that even events classified as sub-bursts should be regarded as independent but physically correlated aftershocks. In this interpretation, all burst frequency behaviors including the sad-trombone effect at short $\dt$ can be explained by a unified model of aftershock activity.

    \item {\bf Extended sub-burst scenario} --The downward drift phenomenon observed among sub-bursts within a single event may extend to time separations as large as $\sim{\cal O}(100)$ ms. Some bursts near this timescale may be actually sub-bursts of longer-duration bursts whose continuum emission among them is undetected due to limited sensitivity. Some FRBs are known to have longer durations (up to $\sim100$ ms) than the more typical short bursts lasting $\lesssim 10$ ms. Suppose the continuum emission of these longer-duration bursts is faint or below the detection threshold. Then, the broader burst may go undetected, leaving only widely separated sub-bursts to be identified as apparent single-component bursts (for a good example, see \citealt{Hewitt2023microshots}, which reports exceptionally bright bursts with faint extended continuum emission lasting $\sim$50 ms, along with multiple burst islands on the order of $\sim$ms durations). In such cases, pairs with $\dt\sim{\cal O}(100)$ ms could consist of sub-bursts from the same event, mixed with independent aftershocks. 
    Our findings, therefore, represent an extension of the sub-burst drift effect, captured over a longer timescale when the low-level continuum emission among bright sub-bursts is too faint to be seen.
    In this scenario, even if the activity duration extends to 100 ms (emitting radio signals below the detection threshold continuously for 100 ms), the sub-pulses would need to have durations of less than 10 ms. This raises questions about their physical origin.
\end{enumerate}

For the first scenario, determining whether the trend of downward frequency shift persists at timescales below 1 ms is beyond the scope of this work, as our sample lacks such extremely short-separation pairs. Still, exploring such temporally-unresolved burst pairs with a sample including extremely bright bursts exhibiting micro-structures \citep[e.g.,][]{Majid2021microshots,Nimmo2022microshots,Hewitt2023microshots} could be an intriguing avenue for future investigation.
The second scenario could be tested with a larger dataset by varying the detection threshold and/or the sampling time resolution, which would also be an interesting direction for future research. 

In summary, our finding suggests that the sad trombone effects can be more broadly interpreted as a statistical manifestation of the aftershock characteristics of repeater FRBs. 
Therefore, the existing theoretical models for the sad trombone effects \citep[e.g.,][]{Wang2019,Metzger2019,Lyutikov2020,Rajabi2020,Tong2022} must be revisited in light of this new discovery. In particular, it needs to explain why these effects persist for longer time separations up to $\dt\sim0.3$ s while their strength diminishes as $\dt$ increases. 

\citet{Totani+23} has shown that there is a statistically significant time correlation between detected bursts, even when they are separated by more than 100 ms, confirming their physical associations. Generally speaking, this physical connection suggests that some mechanism linked to this association shifts the peak frequency of the aftershock bursts to lower values.  

One potential explanation is that  the mainshock induces changes in the surrounding environment (e.g., plasma ejection), and the aftershock occurs while these environmental changes are still present, leading to a frequency shift. The observed stronger correlation between $\dt$ and $\dnup$ for shorter delays could then be attributed to the persistence of these environmental changes over shorter $\dt$. Drawing an analogy to earthquakes, if the observed correlation originates from starquakes in the neutron star's crust, the initial burst could trigger a rupture and alter the surrounding conditions. These changes might then influence subsequent aftershocks occurring in its proximity.

The maximum timescale beyond which the correlation vanishes could also be related to the geometry of the neutron star. For instance, if FRBs are beamed from within localized regions of the neutron star magnetosphere, repeated bursts would be observed only when the observer's line of sight passes through the emission cone. The angular size of this region could set an upper limit on the timescale for correlated bursts.

\subsection{Potential Caveats}

Although not the primary focus of this paper, it is worth mentioning that variations in dispersion measure (DM) can affect the apparent frequency drift or spectral peak position in FRB dynamic spectra.  Larger DM values cause de-dispersed spectra to rotate clockwise on the time-frequency plane (and vice versa), potentially reversing the arrival order of closely separated sub-bursts and altering the sign and/or the absolute value of $\Delta \nu$ for such burst pairs. For bursts with complex structures, determining a consistent DM standard is challenging (Z22). As a result, DM measurement approaches vary across studies: X22, Z23, and J23 determine DM for individual bursts, while Z22 averages DM from well-fit bursts over a day and applies it uniformly. DM measurement methods themselves may also differ, depending on whether they optimize burst structure or signal-to-noise ratio (e.g., \citealt{Hessels2019}; J23). Such potential inconsistencies may introduce systematic differences in frequency measurements, possibly affecting analyses of frequency drifts on the shortest timescales within the sub-burst regime where the sad trombone effect is observed. 

Despite these potential uncertainties in DM, the sad trombone effect appears intrinsic to repeating FRB sources, as supported by detailed dynamic spectrum inspections (see, e.g.,  J23 and Z22 for discussion). Moreover, we find statistical evidence for downward frequency drifts, including those in the sad trombone regime, which persists across datasets analyzed with differing DM optimization standards. This consistency in our result suggests that aforementioned DM-related uncertainties are unlikely to change our conclusions. Future analyses of peak frequency correlation functions, particularly for burst pairs with separations of less than 1 ms, could benefit from careful consideration of DM estimation to minimize potential biases.

\section{Conclusion}
\label{s:conclusion}

In this study, we have analyzed over 4,000 bursts from three of the most active repeating FRB sources: FRB 20121102A, FRB 20201124A, and FRB 20220912A. Our investigation focused on the correlation between peak frequency and burst occurrence times, revealing a universal dependence of the correlation function on peak frequency shifts at short-time separations. The derived correlation function, $\xi(\dnup)$, demonstrates an asymmetric shape with respect to frequency shift, showing a decrease with increasing $\dnup$ from negative to positive values.
This suggests that correlated aftershocks tend to exhibit lower peak frequencies than mainshocks, with this tendency becoming more pronounced at shorter $\dt$.
Through statistical analysis, we found significant evidence for this asymmetry, with the disparity moment of the correlation function yielding values between 1.6--8.9 $\sigma$ for each dataset, and the overall significance reaching 12 $\sigma$ when combining all datasets. 

Our findings also lead to the discovery that the “sad trombone effect”—the downward frequency drift observed in sub-pulses within a single event—not only extends beyond individual sub-bursts but also emerges as a statistical trend across independent aftershocks. While both downward and upward frequency shifts occur, there is a distinct statistical preference for downward shifts at shorter $\dt$. We showed that this behavior spans a broader range of timescales than previously thought. These aftershocks, though physically correlated with the main burst, can occur up to approximately 0.3 seconds after the main burst. This extension of the sad trombone effect into aftershocks indicates that the phenomenon is not isolated to a single burst event, but rather a statistical manifestation of the aftershocks inherent in repeater FRBs.

In conclusion, our results revealed that repeater FRBs exhibit a universal time-frequency correlation structure among repeating sources, offering new insights into the physical processes (independent yet physically correlated aftershocks) governing FRB production. This discovery opens the door to further investigation into the relationship between burst properties and their underlying production mechanisms. In particular, existing theoretical models for the sad trombone effect, which are mainly tuned to explain drifts at short $\dt$, should be re-evaluated in light of our findings. Alternatively, entirely new theoretical frameworks may need to be developed.

Some repeaters show narrow  $\nup$ distributions, while others operate across a wider frequency range, often with activity clustered around specific frequencies \citep{Lyu2024}. In this work, only bursts at GHz frequencies (L band) have been analyzed. Investigating whether the time-frequency correlation varies across different frequency bands would be an interesting avenue for future research. Although current limitations in burst statistics above or below the $\sim$GHz range make this challenging, such studies could provide valuable insights into the aftershock nature of these events and the underlying radiation mechanisms.

\vspace{5mm}

We thank Heng Xu (X22), Bojun Wang (X22), Kejia Lee (X22), Dejiang Zhou (Z22), Yong-Kun Zhang (Z23), and Joscha Jahns (J23 \& J23g) for providing the full numerical data of their FRB catalog and/or relevant information. SY thanks Tetsuya Hashimoto, Jason Hessels, and Tomotsugu Goto for their useful comments. We thank the anonymous referee for providing valuable feedback. This work was initiated at ``Localization of Fast Radio Bursts in Taiwan 2024 (FRB Taiwan 2024)'' at National Yilan University. SY acknowledges the support from NSTC through grant numbers 113-2112-M-005-007-MY3 and 113-2811-M-005-006-. TT was supported by the JSPS/MEXT KAKENHI Grant Number 18K03692.

\vspace{5mm}



\bibliographystyle{aasjournal}

\end{document}